\documentclass[a4paper,UKenglish]{lipics-v2016}

\usepackage{microtype}
\usepackage{multirow}
\usepackage{amssymb}
\usepackage{graphicx}
\usepackage{xspace}
\usepackage[table]{xcolor}
\usepackage{multirow}
\usepackage[titletoc,toc,page]{appendix}
\usepackage{float}
\usepackage{siunitx}
\usepackage{color,soul}
\usepackage{url}

\newcommand{\T}{\ensuremath{\mathcal{T}}\xspace}
\newcommand{\B}{\ensuremath{\mathcal{B}}\xspace}
\newcommand{\Z}{\ensuremath{\mathcal{Z}}\xspace}
\newcommand{\C}{\ensuremath{\mathcal{C}}\xspace}
\newcommand{\F}{\ensuremath{\mathcal{F}}\xspace}

\newcommand{\transfer}{\ensuremath{trans\!f\!er}\xspace}

\newcommand{\dynTM}{DTM\xspace}
\newcommand{\multiDynTM}{MDTM\xspace}
\newcommand{\hide}[1]{\relax}

\newcommand{\redTEEngQuery}{\texttt{TE-QH}\xspace}
\newcommand{\redTEQueryALT}{\texttt{TE-QH-ALT}\xspace}
\newcommand{\redTEUpdate}{\texttt{TE-UH}\xspace}

\newcommand{\dynTMEngQuery}{\texttt{DTM-QH}\xspace}
\newcommand{\dynTMQueryALT}{\texttt{DTM-QH-ALT}\xspace}
\newcommand{\dynTMUpdate}{\texttt{DTM-U}\xspace}
\newcommand{\multiDynTMEngQuery}{\texttt{MDTM-QH}\xspace}
\newcommand{\multiDynTMQueryALT}{\texttt{MDTM-QH-ALT}\xspace}
\newcommand{\multiCriModDynTMEngQuery}{\texttt{McMDTM-QH}\xspace}
\newcommand{\multiCriModDynTMEngQueryALT}{\texttt{McMDTM-QH-ALT}\xspace}
\newcommand{\multiDynTMUpdate}{\texttt{MDTM-U}\xspace}

\title{Multimodal Dynamic Journey Planning
\footnote{This work was partially supported by the Hellenic Foundation for Research and
Innovation, and the General Secretariat for Research and Technology of Greece.}}
\titlerunning{Multimodal Dynamic Journey Planning}
\author[1,2]{Kalliopi Giannakopoulou}
\author[2]{Andreas Paraskevopoulos}
\author[1,2]{Christos Zaroliagis}

\authorrunning{K.~Giannakopoulou, A.~Paraskevopoulos, and C.~Zaroliagis}

\affil[1]{
Computer Technology Institute and Press ``Diophantus'', 26504 Patras, Greece}

\affil[2]{Dept of Computer Eng.~\& Informatics, Univ.~of Patras, 26504 Patras, Greece\\
\texttt{\{gianakok, paraskevop, zaro\}@ceid.upatras.gr }}

\Copyright{K.~Giannakopoulou, A.~Paraskevopoulos, and C.~Zaroliagis}%mandatory, please use full first names. LIPIcs license is "CC-BY";  http://creativecommons.org/licenses/by/3.0/

% mandatory: Please choose ACM 1998 classifications from http://www.acm.org/about/class/ccs98-html . E.g., cite as "F.1.1 Models of Computation".


\EventEditors{}
\EventNoEds{2}
\EventLongTitle{}
\EventShortTitle{16.04.2018}
\EventAcronym{}
\EventYear{2018}
\EventDate{April 16, 2018}
\EventLocation{}
\EventLogo{}
\SeriesVolume{~}
\ArticleNo{XX}
\begin{document}

\maketitle

\begin{abstract}
We present \emph{multimodal DTM}, a new model for multimodal journey planning in public (schedule-based) transport networks.
Multimodal DTM constitutes an extension of the dynamic timetable model (DTM), developed originally for unimodal journey planning.
Multimodal DTM exhibits a very fast query algorithm, meeting the request for real-time response to best journey queries
and an extremely fast update algorithm for updating the timetable information in case of delays. In particular,
an experimental study on real-world metropolitan networks demonstrates that our methods compare favorably with
other state-of-the-art approaches when public transport along with unrestricted w.r.t.~departing time traveling
(walking and electric vehicles) is considered.
\end{abstract}

\section{Introduction}

Journey planning in schedule-based public transport is a most frequent problem nowadays and
journey planners (web or mobile applications) are abundant.
Given as input a timetable associated with a public transportation system, the \emph{journey planning}
problem asks for efficiently answering queries of the form:  ``What is the best
journey from some station $A$ to some other station $B$, provided that I wish to
depart at time $t$?''. In a multimodal setting, the aforementioned problem is
considered in combination with the different modes of transport (train, bus, tram,
walking, EVs, bicycles, car, etc) one can consider. In \emph{multimodal} journey planning the
best route should be provided by a holistic algorithmic approach that not only considers each
individual mode, but also optimizes their choice and sequence.

Depending on the considered metrics and modeling assumptions, the (uni- or multi-modal) journey
planning problem can be specialized into various optimization problems.
In the \emph{earliest arrival-time problem} (EAP), one is interested in finding the
best (or optimal) journey that minimizes the traveling time required to complete it.
In the \emph{minimum number of transfers problem} (MNTP), one is interested in computing
a best journey that minimizes the number of times a passenger
needs to change vehicle during the journey.
Sometimes, these two optimization criteria are considered in combination, giving
rise to multicriteria problems.
We refer to \cite{BDGMPSWW16} for a comprehensive overview on unimodal and multimodal journey planning.

A typical approach to deal with unimodal and multimodal journey planning optimization problems
is to create, in a preprocessing phase, a structure that represents the timetable information,
which subsequently allows for fast answering of queries. There is vast literature on
this approach; see e.g., \cite{BDGMPSWW16} and the references therein.

Journey planning is quite challenging (despite its simple formulation), much more than its route
planning counterpart in road networks. Schedule-based transportation systems exhibit
an inherent time-dependent component that requires more complex modeling assumptions in
order to obtain meaningful results, especially when transfer times from one vehicle to
another has to be taken into account \cite{PSWZ08}.

One additional challenge is to accommodate delays of public transport vehicles that often
occur. The key issue is how to efficiently update the underlying timetable
information structure so that best journey (typically EAP) queries are still answered fast
and optimally with respect to the updated timetable.

The necessity for solving the aforementioned challenges as efficient as possible is crucial,
since otherwise the real-time response requests posed to an actual journey planner, which is
typically in heavy demand (e.g., the one of German railways, may receive in peak hours more
than 420 queries per second), cannot be met both for queries and for digesting delays of schedule-based
vehicles that occur frequently.

There are three main approaches of past work for solving the multimodal journey planning problem \cite{BDGMPSWW16}.
One considers a combined cost function of travel time with penalties for modal transfers (see e.g., \cite{AZC2007,AW2012,MS98}).
Another approach uses the label-constrained shortest path problem to obtain journeys that explicitly include (or
exclude) certain sequences of transportation modes (see e.g, \cite{DPW2012,DDPWW2013,D2016}. 
A third approach considers the computation of Pareto sets of multimodal journeys using a carefully 
chosen set of optimization criteria that aims to provide diverse (regarding the transportation modes) 
alternative journeys (see e.g, \cite{BBS2013,DDPWW2013}. 
Some of these approaches consider also car driving and flights (which are beyond the scope of this paper).

In this work, we consider \emph{urban} multimodal schedule-based public transport (train, bus, tram) along
with unrestricted w.r.t.~departing time traveling (walking and electric vehicles -- EVs). The most
closely related work to ours is that in \cite{DDPWW2013,D2016}, which computes multicriteria multimodal journeys.
Our aim here is to investigate whether the recently introduced unimodal dynamic timetable
model (\dynTM) \cite{CDDFGPZ17}, which handles efficiently EAP journey planning queries and updates extremely
fast the underlying timetable information structure in case of delays, can be extended in the multimodal setting
and can provide competitive query and update times w.r.t.~state-of-the-art approaches.

In this work, we present \emph{Multimodal DTM} (MDTM), an extension of the dynamic timetable model (DTM) \cite{CDDFGPZ17},
which can indeed model urban multimodal journeys while simultaneously offering competitive to state-or-the-art query times
for computing best journeys as well as extremely fast update time for updating the timetable information in case of delays.
We conducted an experimental study on two metropolitan public transport networks (Berlin and London). Our query algorithms
answer multimodal EAP and multicriteria queries very efficiently when public transport (train, bus, tram)
along with unrestricted w.r.t.~departing time traveling (walking and EVs) is considered,
and remain competitive to state-of-the-art approaches even in the case of unlimited walking.
Our timetable information structure can be updated in less than 0.14 milliseconds in case of delays.

The rest of this paper is organized as follows. In Section~\ref{sec:pre}, we present preliminary notions regarding timetable
information modeling and (multimodal) journey planning, along with a succinct review of \dynTM. In Section~\ref{sec:mdtm}
we present the multimodal DTM along with its query and update algorithms. In Section~\ref{sec:exper}, we present our
experimental study. We conclude in Section~\ref{sec:conclusions}.

\section{Preliminaries}
\label{sec:pre}

Schedule-based transportation is described by timetables that determine
the (scheduled) departure and arrival times of public vehicles.
We consider a \emph{timetable} as a set tuple $\T=(\Z,\B,\C)$, where \B is the set of
\textit{stops} (or stations) in which the passengers may embark/disembark on/from a vehicle,
\Z is the set of \textit{vehicles} (train, bus, metro, and any other means of transport
that performs scheduled routes), and \C is the set of \emph{elementary connections} $c = (z, S_d, S_a, t_d, t_a)$,
which represents the travel of a vehicle $z\in\Z$, leaving from stop $S_d\in\B$ at time $t_d$ and
arriving at the immediately next stop $S_a\in\B$ at time $t_a$. Elementary connections of schedule-based
transport are restricted w.r.t.~(the scheduled) departure of the vehicles.

One of the most common models for representing a timetable is the \emph{realistic time-expanded model}
(TE-real) \cite{PSWZ08}. This model encodes a timetable $\T$ into a directed graph $G=(V,E)$ with appropriate
arc weights. In TE-real, nodes represent time events (arrival or departure times of a vehicle at a stop), while
arcs represent either elementary connections (travel of a vehicle between consecutive stops), or transfer
between different vehicles at the same stop, or waiting time between two time events (at the same stop).
The arc weight is the time difference between the time events associated with the endpoints of the arc.
Let us stress that \textit{transfer times} \F introduce realistic transfer restrictions between vehicles,
and represent the required minimum time $\transfer(S)$ that a passenger needs to be transferred between
different vehicles within the same stop $S$.

A reduced version of the model (TE-red), eliminating nodes representing transfer
events (without losing correctness) was also presented in \cite{CDDFGPZ17,PSWZ08}.

\subsection{The Dynamic Timetable Model}

The \emph{dynamic timetable model} (\dynTM) is a new model introduced in \cite{CDDFGPZ17},
aiming at efficiently updating the timetable after a delay of a vehicle.

Given a timetable $\T=(\Z,\B,\C)$, the directed graph $G=(V,E)$ representing \dynTM, is defined as follows:
(1) for each stop $S$ in $\B$, a \emph{switch node} $\sigma_S$ is added to $V$,
representing an arrival or start time event of a traveler at stop $S$;
(2) for each elementary connection $c=(Z,S_d,S_a,t_d,t_a)\in \C$
a \emph{departure node} $d_c$  is added to $V$, and a \emph{connection arc} $(d_c,\sigma_{S_a})$,
connecting $d_c$ to the switch node $\sigma_{S_a}$ of (the immediately next stop) $S_a$, is added to $E$;
(3) for each elementary connection $c=(Z,S_d,S_a,t_d,t_a)\in \C$,
a \emph{switch arc} arc $(\sigma_{S_d},d_c)$, connecting the switch node $s_{S_d}$
of the departure stop $S_d$ to the departure node $d_c$ of $c$ at $S_d$, is added to $E$;
(4) for each vehicle $Z\in\Z$ which travels through the itinerary
$(c_1,c_2,\ldots,c_k)$, an arc (\emph{vehicle arc}), connecting the
departure node $d_{c_i}$ of $c_i$ with the departure node $d_{c_{i+1}}$ of $c_{i+1}$, is
added to $E$, for each $i=1,2,\ldots,k-1$.

The timetable routes are periodic, with period $T_p$ (typically $T_p=1440$). Any transfer and travel time
is assumed to last less than $T_p$. Given two time instances $t_1$ and $t_2$, such that $t_1 \leq t_2$,
$\Delta(t_1,t_2) = t_2-t_1(\bmod~T_p)$ denotes the (cyclic) time difference between them.

The associated time references $t:V\rightarrow \mathbb{R}_{\geq0}$ and the weight function $w:E
\rightarrow \mathbb{R}_{\geq0}$ are defined as follows. The time point $t(v) \in [0,T_p)$ of a departure
node $v \in V$ is fixed and it denotes the scheduled departure time of the associated public
transportation vehicle. The time point $t(v) \in [0,T_p)$  of a switch node $v \in V$ of stop $S
\in \B$, varies and represents any possible start or arrival time at the stop $S$. The weight of each
non switch (i.e., connection and vehicle) arc $e=(u,v) \in E$ is fixed and is set to $w(e)=\Delta(t(u),t(v))$.
The weight of the rest (i.e., switch) arcs $e \in E$ varies and its default value is infinity.

\begin{figure}[htb]
\centering
\includegraphics[height=6cm]{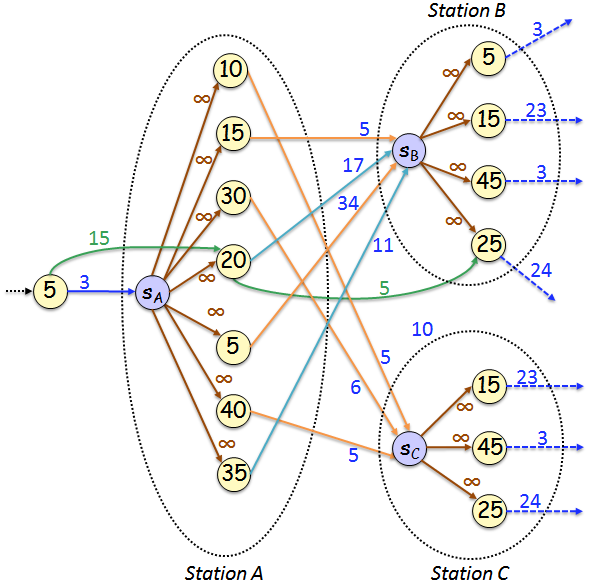}
\caption{A \dynTM graph. Switch nodes are drawn in blue. Departure nodes
(yellow) are associated with the departure time of their corresponding elementary
connection, and are ordered by arrival time at the (arrival) station. Switch arcs are drawn in brown,
while vehicle arcs are drawn in green.}
\label{fig:dtm}
\end{figure}

For each connection $c=(Z,S_d,S_a,t_d,t_a)$, $t_a(d_c)$ or $t_a(c)$ denotes
the arrival time $t_d + w(d_c,\sigma_{S_a})$ at stop $S_a$, departing from $S_d$
via the departure node $d_c$, at time $t(d_c)=t_d$. For each stop, the departure nodes
are ordered by their $t_a(d_c)$ (arrival times at their arrival stop $S_a$).
Moreover, for each switch node $\sigma_S$, we store the stop $S$ it is
associated with, while for each departure node $d_c$, we maintain both the departure
time reference $t_d(c)$ and the vehicle $Z(c)$ of connection $c$ which $d_c$ is associated
with. Figure~\ref{fig:dtm} shows a \dynTM graph. Departures of station $A$ labeled 20 and 35 concern
train connections, while the rest concern bus connections.

\subsection{Multimodal Journey}

A \emph{multimodal transport network} consists of schedule-based public
transport along with road and pedestrian path networks, for supporting
traveling with both unrestricted departure (e.g., for walking, cycling, and driving)
and restricted departure (for embarking on public transport vehicles that follow scheduled
timetables). In contrast to a restricted-departure timetable elementary connection, an
unrestricted-departure connection $(\sigma_{S_A},\sigma_{S_B})$ is defined as an arc
representing a time-independent traveling path from stop $S_A$ to stop $S_B$.

A \emph{multimodal itinerary} is a sequence of trip-paths consisting of unrestricted and
restricted-departure connections $P=(c_1,c_2,\ldots,c_k)$ such that, for each
$i=2,3,\ldots,k$, $S_a(c_{i-1}) = S_d(c_{i})$ and

$$
\Delta(t_a(c_{i-1}),t_d(c_i)) \geq
\left\{
\begin{array}{ll}
0&\mbox{~~if } Z(c_{i-1}) = Z(c_{i})\\ \transfer(S_a(c_{i-1}))
&\mbox{~~otherwise.}
\end{array} \right.
$$

A \emph{multimodal journey query} is defined by a tuple $(S,T,t_o,M)$ where $S \in \B$ is a
departure stop, $T\in \B$ is an arrival stop, $t_o$ is a minimum departure time from $S$,
and $M$ represents the desired transport mode(s). There are two natural optimization criteria used to
answer a timetable query. They consist in finding a multimodal itinerary from $S$ to
$T$ starting (from $S$) at a time after $t_o$ and arriving at $T$ either with the minimum
possible arrival time or with the minimum number of vehicle transfers. These two criteria
define the following core optimization problems: (a) the \emph{Earliest Arrival Problem (EAP)}
is the problem of finding a multimodal itinerary from $S$ to $T$ starting at a time after $t_o$
and arriving at stop $T$ as early as possible, (b) the \emph{Minimum Number of Transfers
Problem (MNTP)} is the problem of finding a multimodal itinerary from $S$ to $T$ starting at
a time after $t_o$ and having as few transfers from one vehicle to another as possible, and
(c) the multicriteria version of EAP and MNTP (\emph{multicriteria multimodal journey query}).

Given a timetable \T, a delay occurring on a connection $c$ is
modelled as an increase of $\delta$ minutes on the arrival time:
$t_a'(c) = t_a(c) + \delta (\bmod~T_p)$. The timetable is then updated according to
some specific policy which depends on the network infrastructure. The new
timetable, called \emph{disposition timetable} $\T'$, differs from $\T$
in the arrival and departure times of the vehicles that depend on $Z(c)$ in \T.

In this work, we consider the simplest policy for handling delays and updating the
timetable (no vehicle waits for a delayed one), which is also considered in similar
works (see e.g., \cite{CDDFGPZ17}). Other policies can be found in e.g.,
\cite{CDDFN09,CDDFNSS09,FSZ09,LSSSP10,SS10}.
Therefore, when a delay occurs on a connection $c$, the only time
references which are updated are those regarding the departure times of $Z(c)$.
Moreover, we assume that the policy does not take into account any possible
slack times and hence the time references are updated by adding $\delta (\bmod~T_p)$.

\section{The Multimodal Dynamic Timetable Model}
\label{sec:mdtm}

In this section, we introduce the \emph{multimodal dynamic timetable model} (\multiDynTM)
aiming at modeling traveling in multimodal transport networks. We also provide the corresponding
algorithms for solving EAP (and MNTP), and for handling delays (updating the timetable).
\multiDynTM is an extension of \dynTM \cite{CDDFGPZ17}. The key difference consists in the
new ordering of the departure nodes within a stop.

Given a timetable $\T=(\Z,\B,\C)$, the directed graph $G=(V,E)$ representing \multiDynTM is defined
similarly to \dynTM, but with the following additional features:

\begin{itemize}
\item For each stop $S \in \B$, its associated departure nodes are grouped in a specific ordering:
(i) a first grouping $\Gamma_1$ is created, where two departure nodes belong to the same group
if the head switch node of their outgoing arcs is identical;
(ii) within each group of $\Gamma_1$, a second grouping $\Gamma_2$ is created, where two departure nodes belong to the same group
if the transport mode they represent is identical (i.e., departures of the same means of transport are grouped together);
and (iii) the departure nodes within each group of $\Gamma_2$ are ordered by increasing arrival time at the head switch nodes
of their outgoing arcs.

\item For each unrestricted-departure connection from stop $S_A$ to stop $S_B$, a switch-switch
arc $(\sigma_{S_A}, \sigma_{S_B})$ is added to $G$.

\end{itemize}

\begin{figure}[htb]
\centering
\includegraphics[height=6cm]{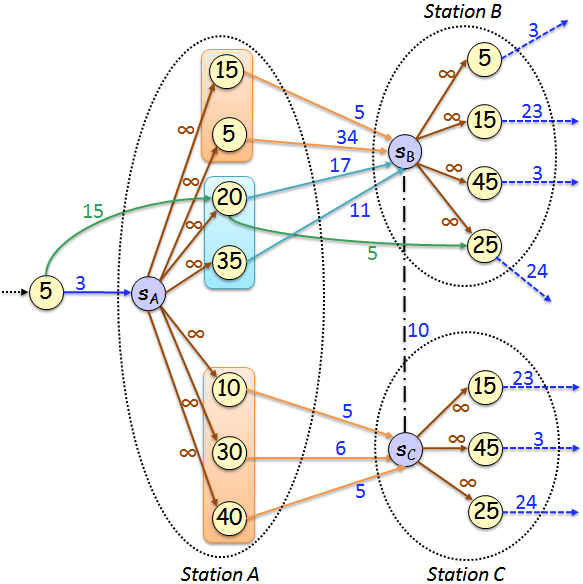}
\caption{The \multiDynTM graph corresponding to the \dynTM graph of Figure~\ref{fig:dtm}. Departure nodes
grouping: light blue (brown) are train (bus) connections.
The switch-switch arc (dotted black) introduces an unrestricted-departure connection between the stops.}
\label{fig:mdtm}
\end{figure}

Let $D_{S_d}(S_a, M)$ denote a group of departure nodes resulted from the aforementioned grouping, having
departure stop $S_d$, arrival stop $S_a$, and transport mode $M$. Figure~\ref{fig:mdtm} shows the
\multiDynTM graph corresponding to the \dynTM graph of Figure~\ref{fig:dtm}. Then, $D_{S_A}(S_B, bus)$
includes departure nodes 5 and 15 of stop $S_A$ that correspond to bus connections departing from $S_A$ and
arriving at $S_B$.

\subsection{Query Algorithm}
\label{sec:query}

We shall now present our query algorithm, named \multiDynTMEngQuery, for solving EAP on a \multiDynTM graph $G$.
An EAP query $(S,T,t_s, M_{choices})$ is answered by executing a modified Dijkstra's algorithm on $G$,
starting from the switch node $\sigma_S$ of stop $S$.

Before discussing the details of our algorithm, we will first describe the additional data
structures used by \multiDynTMEngQuery with respect to the classic Dijkstra's algorithm.
In particular, for each switch node of a stop, algorithm \multiDynTMEngQuery maintains a set
of \emph{earliest arrival index tables}, whose construction and contents are described below.

Each departure node $d$ is associated with a
departure time $t_d(d)$ and an arrival time $t_a(d)$. In general, all the departure nodes with the same
departure stop $S_d$ can be ordered by either departure time or arrival time at an arrival stop. The first
ordering favors an efficient search on the departure nodes to get the valid
routes which have a valid departure time, i.e., a departure time greater than or equal to $t(\sigma_{S_d})$ at
which the traveler is at $S_d$. The second ordering favors an efficient search on the departure nodes on the
optimal routes which provide the earliest arrival times to their adjacent stops. The second ordering is
already applied within the existing ($\Gamma_1$ and $\Gamma_2$) departure groups of $G$.
However, we would also like to have the advantage of the first ordering. Therefore, for the purpose of searching
effectively the departure nodes that have both valid departure times from stop $S_d$ and earliest
arrival times to an adjacent arrival stop $S_a$, we introduce the \emph{earliest arrival index}
tables. Let $I_{S_d}(S_a, M)$ denote such a table, consisting of departure nodes with
departure stop $S_d$, arrival stop $S_a$ and transport mode $M$. $I_{S_d}(S_a, M)$ is
constructed as follows: Let $d_1, d_2, .., d_k \in D_{S_d}(S_a, M)$ be the sequence of the departure nodes,
ordered by arrival time at $S_a$, for a trip departing from stop $S_d$
and arriving to stop $S_a$ with transport mode $M$. Initially, $I_{S_d}(S_a, M)$ is empty.
Node $d_1$ is inserted in $I_{S_d}(S_a, M)$ and $t_{max}=t(d_1)$
is the current max departure time. Afterwards, for $i=2,...,k$, if $t_{max} < t(d_i)$, then $t_{max} =
t(d_i)$ and $d_i$ is inserted at the end of $I_{S_d}(S_a, M)$ table; otherwise, $d_i$ is skipped. If the
table contains the departure nodes $v_1, ... , v_l$, $l \leq k$, and for some $i$, $t(\sigma_{S_d}) \in [t(v_i), t(v_{i+1}))$,
then $v_i$ is the first departure node to start the search of the earliest arrival time at $S_a$.
This allows us to bypass the departure nodes, before $v_i$ in the arrival ordered sequence, with departure time less than $t(v_i)$.

Table~\ref{table:eaindex} shows the contents of table $I_{S_A}(S_B, M=\{bus, train\})$, for the example shown
in Figure~\ref{fig:mdtm}. If a traveler has arrived or has started its journey from stop $S_A$ at time
$t(s_A)=25 > 20$, then to continue at the adjacent stop $S_B$, the search of valid and optimal path-solutions can start after the departure node 20.

\begin{table}[htb]
\centering
\resizebox{.3\columnwidth}{!}{%
\begin{tabular}{||c | c | c||}
\hline
depNode & depTime & arrTime \\ [0.5ex]
\hline\hline
$d_{15}$ & 15 & 20  \\
\hline
$d_{20}$ & 20 & 37  \\
\hline
$d_{35}$ & 35 & 46  \\
\hline
\end{tabular}}
\caption{An earliest arrival index example. $I_{S_A}(S_B, M=\{bus, train\})$ includes the most important departure nodes, in terms of valid departure and earliest arrival time, with departure stop $S_A$, arrival stop $S_B$ and traveling with bus or train.}
\label{table:eaindex}
\end{table}

To reduce the size and the operations in the priority queue of the query
algorithm, we insert in it only the switch nodes and change the arc relaxation as described below
(iteration step). The \multiDynTMEngQuery algorithm works as follows.

\emph{Initialization}. The switch node $\sigma_{S_o}$ of the origin stop $S_o$ is inserted in the priority queue,
with distance $dist[\sigma_{S_o}]=t_s$ and time $t(\sigma_{S_o})=t_s$. During the algorithm execution, provided that the traveler is already
at $S_o$ at time $t_s$ the minimum transfer time of $S_o$ is set to $transfer(S_o)=0$.

\emph{Iteration}. At each step, a switch node $\sigma_{S_x}$ is extracted from the priority queue.
The node $\sigma_{S_x}$ is settled having got the earliest arrival time for the optimal journey departing from $S_o$ at time
$t_s$ and arriving to $\sigma_{S_x}$ at time $t(\sigma_{S_x})=dist[\sigma_{S_x}](\bmod~T_p)$. Then the algorithm relaxes the outgoing arcs of $\sigma_{S_x}$
following a node filtering and blocking process:
\begin{itemize}
\item[(i)] Using the existing grouping of departure nodes ($\Gamma_1$ and $\Gamma_2$), the departure node groups corresponding
to non-selected transportation modes are skipped;

\item[(ii)] Using the earliest arrival index tables, the algorithm can skip the departure nodes of stop $S_x$
that have an earlier than $dist[\sigma_{S_x}]$ departure time or they provide non-optimal arrival times to the next adjacent stops.
In particular, after $\sigma_{S_x}$ is extracted, then for each adjacent arrival stop $S_a$ of stop $S_x$ and
for each enabled transport mode $M \in M_{choices}$: (1) a binary search is performed on the index table $I_{S_x}(S_a, M)$
for getting the first contented departure node $d_r$ with $t(d_r) > t(\sigma_{S_x})$
that provide the earliest arrival time at stop $S_a$; and (2) an arc relaxation step
is performed based on those departure nodes.

\end{itemize}

Let the sequence of the outgoing switch arcs of $\sigma_{S_x}$ be $e_1, e_2, ..., e_{r-1} , e_r, ... , e_k$,  which corresponds to a travel
using the transport mode $M$ and arriving at the stop $S_a$. Let $e_i=(\sigma_{S_x}, d_i)$, $i=1,...k$, and let the node $d_r$ be the departure
node that is returned by  $I_{S_x}(S_a, M)$. Within the current time period $T_p$, arcs $e_1, e_2, ... , e_r$ can be safely skipped, because
they provide earlier departures or non-optimal arrival paths from $S_x$ to $S_a$. The first arc that is relaxed is $e_r$. Provided that the
next switch arcs and their departure node heads are ordered by arrival time, the algorithm relaxes the arcs $e_r, ..., e_k, e_1, ... , e_{r-1}$
and it stops as soon as falls over a departure node $d_i$ with
(a) $\Delta(t(\sigma_{S_x}), t(d_i)) > transfer(S_x)$ and
(b) $dist[d_i] + w(d_i, \sigma_{S_a}) > dist[\sigma_{S_a}]+ transfer(S_a)$.
The  first condition ensures the minimum transfer time for the traveler in using a different vehicle to
continue his/her travel. The second condition ensures that we will not miss optimal paths in ${S_a}$ with no transfer.

At each case, if the departure node head $d_i$ of the switch arc $e_i$ is visited for the first time or it has from a
previous step a greater  distance, then we set $dist[d_i] = dist[\sigma_{S_x}]+\Delta(t(\sigma_{S_x}), t(d_i))$ and
$w(\sigma_{S_x}, d_i) = dist[d_i]-dist[\sigma_{S_x}]$. When the distance is updated, the algorithm relaxes also the
outgoing arcs of the departure node $d_i$. For its outgoing arc $(d_i, \sigma_{S_a})$, if $\sigma_{S_a}$ is
visited for the first time or it has from a previous step a greater distance, then we set $dist[\sigma_{S_a}] = dist[d_i]+w(d_i,\sigma_{S_a})$.
Also if there is a vehicle (departure-departure) arc $(d_i, d_j)$, then for the associated vehicle we also relax the
outgoing arcs of the departure nodes $d_j,..., d_{t}$ at the next stops at which the vehicle passes, for the same route.
In that case, we can stop if we fall over a departure which has an outgoing arc to a switch node which has not yet been
visited or extracted from the priority queue. The algorithm finishes when the switch node $\sigma_{S_T}$
of the destination stop $S_T$ is settled.

\subsection{Improved Query Algorithm}

In order to boost the performance of the query algorithm \multiDynTMEngQuery,
we adapt the ALT heuristic \cite{GH05}. The combined algorithm is named \multiDynTMQueryALT.
ALT is a goal-directed technique, i.e., its
main aim is that of pushing the shortest-path search faster towards the target stop $t$.
This is achieved by adding a \emph{feasible potential} to the priority of each node in the queue.
The feasible potentials are computed as follows. Given a set of nodes
$L\subseteq V$ called \emph{landmarks}, the feasible potential of a node $u\in
V$ towards a target $t$ is computed as $\pi_t(u) = \max_{\ell\in
L}\max\{dist(u,\ell) - dist(t,\ell); dist(\ell,t)-dist(\ell,u)\}$. By the triangle
inequality, it follows that $\pi_t(u)$ is a lower bound to the distance $dist(u,t)$ and
this is enough to prove the correctness of the shortest path algorithm
(see~\cite{GH05} for more details). It is easy to see that the tighter the lower
bounds are, the more narrowed the search space becomes (i.e., the faster the query
algorithm performs). Therefore, choosing good landmarks that provide tight lower
bounds is a fundamental part of the preprocessing phase of ALT.

As in \dynTM \cite{CDDFGPZ17},  we apply in \multiDynTM an approach similar to that
proposed in~\cite{DPW09}. In particular, we select as landmarks the switch nodes,
each of which represents the arrival node group of a station. Therefore the lower bound distance,
$dist(s_{A}, s_{B})$, between two switch nodes, $s_{A}$ and $s_{B}$, denotes the
minimum travel time among connections traveling from station $A$ to station $B$.
These lower bound distances can be computed during a preprocessing phase by running
single-source queries from each switch node. The
tightest lower bounds can be obtained by storing all pair
station distances $O(|\B|^2)$ and by computing all-pairs shortest paths on the
condensed version of the input graph (see~\cite{DPW09}). This makes sense
particularly when the stations are relatively few in number.

We combined ALT along with our query algorithm in Section~\ref{sec:query}. This combination
reduces considerably the search space and leads to a more efficient algorithm.

\subsection{The Multicriteria Multimodal Query Algorithm}

In order to provide best journeys for a vector of cost functions over the multimodal transport options, we introduce
the \emph{multicriteria} extensions of \multiDynTMEngQuery and \multiDynTMQueryALT.
In that case, in addition to computing journeys with a variety of transport modes,
the optimal Pareto set of journeys is computed on the EA and MNT criteria
(a set of pairwise non-dominating journeys, each of them being better to at least one objective criterion and no worse in all other criteria).
Since all Pareto-optimal journeys are exponential in number, we focus on finding a solution that minimizes MNT, while retaining the EA below
a given threshold $P$ (a variant also considered in \cite{PSWZ08}).
Our multicriteria algorithm \multiCriModDynTMEngQuery is described below. Its combination with ALT
will be called \multiCriModDynTMEngQueryALT.

Let $(S,T,t_s, M_{choices})$ be a multicriteria (EA, MNT) query, starting from the switch node $\sigma_S$ of stop $S$.
The number of transfers is taken into account by setting the weight of all switch-departure arcs to 1
(representing a transfer between vehicles) and the weight of the rest of the arcs to 0. Due to the
modeling, every single switch node in \multiDynTM can have at least as
many Pareto-optimal solutions as its incoming arcs. Initially, the cost minimization is on EA.
Therefore, when the target switch node is settled, we have found the first (EA,MNT) Pareto optimal journey,
with the minimum arrival time $A_{min}$. We then let Dijkstra's algorithm continue; whenever the target switch node
is explored again with a smaller number of transfers than in any of the already found Pareto-optimal solutions,
a new Pareto-optimal journey is found. The algorithm stops when all journey solutions, with arrival time to the
target stop less or equal than $P\cdot A_{min}$, have been found.

\subsection{Update Algorithm}
\label{sec:dynTM-update}

We shall now present our update algorithm, based on \cite{CDDFGPZ17}, named \multiDynTMUpdate, for
updating the timetable when a delay occurs, that is, for updating the
corresponding \multiDynTM graph.

Given a timetable \T, we assume that a delay $\delta$ occurs first on a
connection $c_{0}$ of \T, and it is propagated to the (affected)
connections $c_{0}, c_{1}, ..., c_{k}$, which are performed  by the same vehicle.
Also let $d_0, d_1, ..., d_k$ be the departure nodes corresponding to the affected
connections.
If \T is represented as a \multiDynTM graph $G$, then the \multiDynTM update algorithm
computes the \multiDynTM graph $G'$, corresponding to the disposition timetable \T$'$,
as follows.
\begin{itemize}

\item Edge weight increase: Starting with $c_{0}=(Z,S_d,S_a,t_d,t_a)$, the weight of arc $(d_{t_d}, \sigma_{S_a})$
is increased by $\delta$.

\item Node reordering: For each of the other connections $c_i$, $i=1,..,k$, its associated
departure node $d_i$ has its departure time $t_d(d_i)$ increased by $\delta$. Due to that increase, the
arrival time ordering of the departure nodes on the affected stops may be invalidated. Hence, along with the
new arrival times, the departure node $d_i$ might need to be moved to its correct position within its group,
i.e., before a departure node with arrival time greater than $t_a(d_i)$.

\end{itemize}

\section{Experimental Evaluation}
\label{sec:exper}

\subsection{Experimental setup}

In this section, we present our experimental study to assess the
performance of the algorithms presented in the previous sections.
The experiments have been performed on a workstation equipped with an Intel
Quad-core i5-2500K 3.30GHz CPU and 32GB of main memory.
All algorithms implemented in \texttt{C++} and compiled with \texttt{gcc} (v4.8.4, optimization level O3).

\subsection{Input Data and Parameters}

The input data for creating the multimodal transport networks are (a) timetable data sets in the General
Transit Feed Specification (GTFS) format, containing various means of public transport and (b) road and
pedestrian network data sets in the Open Street Map (OSM) format. The integrated networks concern the
metropolitan areas of Berlin and London. The source of the timetable data for London is \cite{tfluk} and
for Berlin is \cite{trfds}. The source of the road and pedestrian network data is \cite{osmfds}.
The packed-memory graph structure \cite{MMPZ13} was used for representing the input instances
in all implemented algorithms.

Tables~\ref{table:input-sizes} and~\ref{table:input-stats} provide detailed information
about the input timetables. In Table~\ref{table:input-sizes}, we report, for each timetable,
the number of stations $|\B|$ and the number of elementary connections $|\C|$ between
stops (a proxy for size), as well as the number of nodes $|$V$|$ and arcs $|$E$|$ of the
corresponding graph, for each model. In the same table we include the sizes of the realistic
time-expanded (TE-real) and the reduced time-expanded models (TE-red) \cite{CDDFGPZ17,PSWZ08} for comparison.

\begin{table}[htb]
\centering
\resizebox{\columnwidth}{!}{%
\begin{tabular}{|c||c|c||c|c||c|c||c|c||c|c|}
\hline
\multirow{2}{*}{\textbf{map}} &
\multirow{2}{*}{\textbf{$|\B|$}} &
\multirow{2}{*}{\textbf{$|\C|$}} &
\multicolumn{2}{c||}{\textbf{TE-real}} &
\multicolumn{2}{c||}{\textbf{TE-red}} &
\multicolumn{2}{c||}{\textbf{\dynTM}} &
\multicolumn{2}{c|}{\textbf{\multiDynTM}} \\
\cline{4-11} &&& \textbf{$|$V$|$} & \textbf{$|$E$|$} & \textbf{$|$V$|$} & \textbf{$|$E$|$} &
\textbf{$|$V$|$} & \textbf{$|$E$|$} & \textbf{$|$V$|$} & \textbf{$|$E$|$} \\
\hline
Berlin & 12\,838 & 4\,322\,549 & 12\,967\,647 & 21\,612\,745 & 8\,645\,098 & 17\,024\,138 & 4\,335\,387 & 12\,701\,695 & 4\,335\,387 & 12\,708\,568 \\
\cline{1-11}
London & 20\,843  & 14\,064\,967 & 42\,194\,901 & 70\,324\,835 & 28\,129\,934  & 55\,758\,468 & 14\,085\,810 & 41\,837\,355 & 14\,085\,810 & 41\,856\,048 \\
\cline{1-11}
\hline
\end{tabular}}
\caption{Tested timetables and sizes of the corresponding graphs.}
\label{table:input-sizes}
\end{table}
\begin{table}[htb]
\centering
\resizebox{.3\columnwidth}{!}{%
\begin{tabular}{|r|c|c|}
\hline
\multicolumn{1}{|c|}{\textbf{\T}} & Berlin & London \\ \hline \hline
transfer time            & 0.7    & 0.8    \\ \hline
adjacent stops           & 2.7    & 1.2    \\ \hline \hline
bus                      & 76\%   & 98\%   \\ \hline
train                    & 15\%   & 2\%    \\  \hline
tram                     & 9\%    &        \\ \hline
\end{tabular}}
\caption{Timetable characteristics: average transfer time (mins), average degree of adjacent stops/stations and percentage of transportation means (\% of total).}
\label{table:input-stats}
\end{table}

\vspace*{-0.5cm}
In Table~\ref{table:input-stats}, we report, for each timetable,
the average transfer time for changing vehicles, the average number of adjacent
stops or stations, and the percentage of the existed transportation means to the total number of
the elementary connections.

For experimenting with \multiDynTM, we additionally added non-restricted departure traveling paths,
using the following two approaches:
\begin{itemize}

\item Limited walking and driving travel time paths on transitively closed pedestrian and road networks.
Via the pedestrian networks, we added single foot-paths for enabling walking between nearby
stops. The foot-paths that were selected and added had shortest travel time of at most 10mins, with
walking speed 1m/sec. Also, via the road networks, we added free flow speed driving-paths for enabling the
driving between stops with EVs. For this scenario, we considered 10 EV-stations providing public
communal EVs with shortest travel time of at most 1 hour. In addition, the driving-paths connect
only EV-stations. In the Berlin instance, the switch-switch arcs representing foot-paths are 2381
and the driving-paths are 39. In the London instance, the switch-switch arcs representing
foot-paths are 37226 and driving-paths are 60.

\item Unlimited walking travel time paths on the full pedestrian network. For this purpose,
we connected each switch node in the public transit network with the nearest node in the pedestrian network
by an access edge. This approach was inspired by that in \cite{WZ2017}. In the Berlin instance, the 
embedded pedestrian network had 932108 nodes and 1059556 edges. In the London instance, the embedded 
pedestrian network had 1520056 nodes and 1653052 edges.
\end{itemize}

To compute efficiently the different nearest node pairs in both cases,
we used the tree data structure \emph{R-tree}\footnote{\emph{R-tree} is a balanced search
tree that can order and group nearby geographical points by their minimum bounding geographical rectangle.
The tree organizes its data in pages, each one of a maximum number of entries, \emph{M}. The nearest neighbor
search can be done efficiently in $O(\log_{M}n)$ time, where $n$ is the number of the geographical points.} \cite{G84}.
The combination of the query algorithms with ALT requires a preprocessing phase, whose requirements
are reported in Table~\ref{table:alt-reqs}.
\begin{table}[htb]
\centering
\begin{tabular}{|r|c|c|}
\hline
\multicolumn{1}{|c|}{\textbf{\T}}
            & Berlin & London\\ \hline
Space (GB)  & 0.6 & 1.4\\ \hline
Time (mins) & 0.73 & 1.79\\ \hline
\end{tabular}
\caption{ALT-based preprocessing time and space requirements for all timetables.}
\label{table:alt-reqs}
\end{table}

\subsection{Experimental Results}

For each input instance, we generated 1000 random queries consisting of source and target stop pairs,
along with a departure time at each source stop. In the experimental evaluation we
have included the EA query algorithms \redTEEngQuery, \redTEQueryALT, for TE-red \cite{CDDFGPZ17},
\dynTMEngQuery\, \dynTMQueryALT, for \dynTM \cite{CDDFGPZ17}, and the new algorithms
\multiDynTMEngQuery\, \multiDynTMQueryALT, for \multiDynTM. For the latter, we have also
included the multicriteria (EA,MNT) query algorithms \multiCriModDynTMEngQuery and \multiCriModDynTMEngQueryALT with
threshold $P$ on EA 100\% and 120\% (denoted with extension \texttt{1.0} and \texttt{1.2}, respectively).
Note that \texttt{QH} denotes an optimized version of Dijkstra's algorithm (with the corresponding node blocking and
extended arc relaxation methods \cite{CDDFGPZ17}) and \texttt{QH-ALT} denotes its ALT extension \cite{CDDFGPZ17}.
The results of the algorithms for answering multimodal queries are reported in Table~\ref{table:queries},
where we include only the ALT-based algorithms (marked with the \texttt{ALT} suffix) that had a much better performance
(for completeness, we report the comparison among algorithms with and without the ALT heuristic  
in Table~\ref{table:queries-alt} in the Appendix).

\begin{table}[htb]
\centering
\small
\hspace*{-2.3cm}
\resizebox{\columnwidth}{!}{%
\begin{tabular}{|c|c|c|c|c|c|c|c|c|c|}
\hline
\multirow{2}{*}{~} & \multirow{2}{*}{Algorithm} & \multirow{2}{*}{MC} & \multicolumn{5}{c|}{Travel Modes} & \multicolumn{2}{c|}{\begin{tabular}[c]{@{}c@{}}Query {[}ms{]}\end{tabular}} \\ \cline{4-10}
  & & & Bus & Train & Walk & EV/Car & Cycle & L-Walk & U-Walk  \\ \hline
\multirow{4}{*}{\rotatebox{90}{Berlin}} & \redTEQueryALT \cite{CDDFGPZ17} & & $\bullet$ & $\bullet$ & & & & 6.88 &  \\ \cline{2-10}
 & \dynTMQueryALT \cite{CDDFGPZ17} & & $\bullet$ & $\bullet$ & & & & 12.17 & \\ \cline{2-10}
 & \multiDynTMQueryALT & & $\bullet$ & $\bullet$ & & & & 6.12 &  \\ \cline{2-10}
 & \multiDynTMQueryALT & & $\bullet$ & $\bullet$ & $\bullet$ & $\bullet$ & & 8.49 & 105.12 \\ \hline
 \multirow{8}{*}{\rotatebox{90}{London}} & \redTEQueryALT \cite{CDDFGPZ17} & & $\bullet$ & $\bullet$ & & & & 5.14 &  \\ \cline{2-10}
 & \dynTMQueryALT \cite{CDDFGPZ17} & & $\bullet$ & $\bullet$ & & & & 10.25 &  \\ \cline{2-10}
 & \multiDynTMQueryALT & & $\bullet$ & $\bullet$ & & & & 4.17 &  \\ \cline{2-10}
 & \multiDynTMQueryALT & & $\bullet$ & $\bullet$ & $\bullet$ & $\bullet$ & & 6.10 & 114.88 \\ \cline{2-10}
 & \multiCriModDynTMEngQueryALT-\texttt{1.0} & $\bullet$ & $\bullet$ & $\bullet$ & $\bullet$ & $\bullet$ & & 6.29 & 216.36 \\ \cline{2-10}
 & \multiCriModDynTMEngQueryALT-\texttt{1.2} & $\bullet$ & $\bullet$ & $\bullet$ & $\bullet$ & $\bullet$ & & 15.44 & 360.94 \\ \cline{2-10}
 & \texttt{MCR-ht} \cite{DDPWW2013,D2016}    & $\bullet$ & $\bullet$ & $\bullet$ & $\bullet$ &  & $\bullet$ &  & 361.23 \\ \cline{2-10}
 & \texttt{MR-}$\infty$\texttt{-t10} \cite{DDPWW2013,D2016} & $\bullet$  & $\bullet$ & $\bullet$ & $\bullet$ &  & $\bullet$ & 21.47 & \\ \hline
\end{tabular}}
\hspace*{-2cm}
\caption{Comparison between query algorithms. L-Walk (U-Walk) denotes a query algorithm with limited (unlimited) walking.
\multiCriModDynTMEngQueryALT-\texttt{1.0} (\multiCriModDynTMEngQueryALT-\texttt{1.2}) denotes \multiCriModDynTMEngQueryALT with $P=1$ ($P=1.2$).
Bullets ($\bullet$) indicate the options taken into account. MC denotes a multicriteria journey
on arrival time and number of transfers (and walking duration for \texttt{MCR-ht}).}
\label{table:queries}
\end{table}

We added in Table~\ref{table:queries} the query times of the best
previous (\emph{RAPTOR} based) approaches \texttt{MCR-ht} and \texttt{MR-}$\infty$\texttt{-t10}\footnote{
\texttt{MCR-ht} weakens the domination rules by trading off walking and arrival time.
In \texttt{MR-}$\infty$\texttt{-t10} the walking duration is not used as criterion and it is limited to 10 minutes.
}
in \cite{DDPWW2013,D2016}. We stress that the former computes
multicriteria (on arrival time, number of transfers, and walking duration)
multimodal journeys, while the latter computes multicriteria (on arrival time and number of transfers)
multimodal journeys. The times are scaled versions of those reported in \cite{DDPWW2013,D2016} using the
benchmark for scaling factors in \cite{timeScalRef}.
Note that since the \texttt{MCR-ht}, \texttt{MR-}$\infty$\texttt{-t10},
\multiCriModDynTMEngQueryALT-\texttt{1.0} and \multiCriModDynTMEngQueryALT-\texttt{1.2}
report multicriteria multimodal queries, it is natural that they take more time than regular
multimodal EAP (unicriterion) query algorithms. This is also true for the case of unlimited walking,
due to the much larger search space explored by the algorithms.
Nevertheless, \multiCriModDynTMEngQueryALT-\texttt{1.0} and \multiCriModDynTMEngQueryALT-\texttt{1.2}
are competitive to \texttt{MCR-ht} and \texttt{MR-}$\infty$\texttt{-t10}.

 From Tables~\ref{table:queries} and \ref{table:queries-alt}, we observe
that \multiDynTMEngQuery achieves a smaller query time than \redTEEngQuery and \dynTMEngQuery.
This is due to the grouping and ordering of nodes within each station.
In the reduced model TE-red, within each stop, the arrival time events are not merged and the departure nodes are
ordered by departure time. Therefore, TE-red has a disadvantage on finding optimal paths between stops
and an advantage on finding valid paths between the stops. In \dynTM, within each stop, the arrival time
events are merged into a switch node and the departure time events are ordered by arrival time (at the next station).
Therefore, \dynTM has an advantage on finding the optimal paths between the stops and an disadvantage on
finding the valid paths between the stops. In comparison to \dynTM, \multiDynTM has the departure nodes
ordered by arrival time within a set of groups ($\Gamma_1$ and $\Gamma_2$), along with the earliest arrival index tables.
In that way, \multiDynTM gains an advantage on blocking non-selected transportation modes and also non-valid paths between
the stops. In all the cases, having extended the optimized Dijkstra's variant query algorithm with the ALT goal-directed speedup
technique leads to a significant decrease in query time, by at least 40\%. With ALT the query algorithms need
less iterations on finding the target stop. Our multicriteria algorithms with $P=1.0$ are faster than those with
$P=1.2$, since they compute less journeys.

For evaluating updates (occurring after a delay), 1000 elementary connections were randomly selected, for each
input instance, and for each elementary connection we randomly generated a delay affecting the corresponding
train or bus, chosen with uniform probability distribution between 1 and 360 minutes.
\begin{table}[htb]
\centering
\begin{tabular}{|c|c|c|c|c|}
\hline
\multirow{2}{*}{Instance} & \multirow{2}{*}{Algorithm} & \multicolumn{2}{c|}{Travel Modes} & \multirow{2}{*}{\begin{tabular}[c]{@{}c@{}}Update\\ {[}$\mu$s{]}\end{tabular}} \\ \cline{3-4}
 & & Bus & Train & \\ \hline
 \multirow{3}{*}{Berlin} & \redTEUpdate \cite{CDDFGPZ17} & $\bullet$ & $\bullet$ & 238.5 \\ \cline{2-5}
 & \dynTMUpdate \cite{CDDFGPZ17} & $\bullet$ & $\bullet$ & 80.2 \\ \cline{2-5}
 & \multiDynTMUpdate & $\bullet$ & $\bullet$ & 84.4 \\ \hline
\multirow{3}{*}{London} & \redTEUpdate \cite{CDDFGPZ17} & $\bullet$ & $\bullet$ & 477.2 \\ \cline{2-5}
 & \dynTMUpdate \cite{CDDFGPZ17} & $\bullet$ & $\bullet$ & 122.8 \\ \cline{2-5}
 & \multiDynTMUpdate & $\bullet$ & $\bullet$ & 137.5 \\ \hline
\end{tabular}
\caption{Comparison among update algorithms for the TE-red, \dynTM and \multiDynTM models.}
\label{table:updates}
\end{table}
In the experimental evaluation we have included the update algorithms
\redTEUpdate for TE-red \cite{CDDFGPZ17}, \dynTMUpdate for \dynTM \cite{CDDFGPZ17},
and the new algorithm \multiDynTMUpdate for \multiDynTM.
The experimental results of the update algorithms are reported in Table~\ref{table:updates}.
The updates times measure the average computational times for updating the graphs when a delay in
a transportation vehicle itinerary has to be absorbed. In \dynTMUpdate, only (at most)
two arc weights and few node time references need to be changed in the original graph to keep
the EAP queries correct. Although \multiDynTMUpdate has an additional computation cost to maintain
the earliest arrival index table for any node time reference update, the time of \multiDynTMUpdate
is competitive to that of \dynTMUpdate. The updating algorithm in both cases are less than 138 $\mu$s.
That is due to the fact that the number of stations where something
changes, as a consequence of a delay, is small with respect to the size of the whole set of stations $|\B|$.

\section{Conclusions and Future Work}
\label{sec:conclusions}

We described the Multimodal DTM, a model for multimodal route planning that constitutes an extension
of the dynamic timetable model (DTM) (originally developed for unimodal journey planning) that compares
favorably with other state-of-the-art multimodal route planners.

We are currently developing a mobile application for multimodal route planning, using Multimodal DTM
as the core routing engine of the cloud-residing component. Our multimodal journey planner can also be
combined with the mobile application developed in \cite{GNPZ2017} that allows users to evaluate the
suggested (by the planner) routes. We are also working in developing multicriteria multimodal journeys
with more (than two) traveling options/criteria.

\medskip

\noindent
\textbf{Acknowledgments.}
The last author is indebted to Tobias Z\"undorf for various interesting discussions.

\bibliographystyle{plainurl}% the recommended bibstyle

\begin{thebibliography}{}

\end{thebibliography}


\begin{thebibliography}{50}

\bibitem{AZC2007}
G.~Aifadopoulou, A.~Ziliaskopoulos, and E.~Chrisohoou. 
\newblock Multiobjective optimum path algorithm for passenger pretrip planning in multimodal transportation networks.
\newblock \emph{J. Transp. Res. Board} 2032(1), 26–34 (2007).

\bibitem{AW2012}
L.~Antsfeld and T.~Walsh. 
\newblock Finding multi-criteria optimal paths in multi-modal public transportation networks using the transit algorithm. 
\newblock In Proceedings of the 19th ITS World Congress (2012).

\bibitem{BBS2013}
H.~Bast, M.~Brodesser, and S.~Storandt.
\newblock Result diversity for multi-modal route planning. 
\newblock In \emph{Algorithmic Approaches for Transportation Modeling, Optimization, and Systems} -- ATMOS 2013, 
OASIcs, pp. 123–136 (2013).

\bibitem{BDGMPSWW16}
H.~Bast, D.~Delling, A.~V. Goldberg, M.~M{\"{u}}ller{-}Hannemann, T.~Pajor,
  P.~Sanders, D.~Wagner, and R.~F. Werneck.
\newblock Route planning in transportation networks.
\newblock In {\em Algorithm Engineering - Selected Results and Surveys}, volume
  9220 of {\em Lecture Notes in Computer Science}, pages 19--80. 2016.

\bibitem{CDDFN09}
S.~Cicerone, G.~D'Angelo, G.~{Di Stefano}, D.~Frigioni, and A.~Navarra.
\newblock Recoverable robust timetabling for single delay: Complexity and
  polynomial algorithms for special cases.
\newblock {\em Journal of Combinatorial Optimization}, 18(3):229--257, 2009.

\bibitem{CDDFNSS09}
S.~Cicerone, G.~D'Angelo, G.~{Di Stefano}, D.~Frigioni, A.~Navarra,
  M.~Schachtebeck, and A.~Sch{\"o}bel.
\newblock Recoverable robustness in shunting and timetabling.
\newblock In {\em Robust and Online Large-Scale Optimization}, volume 5868 of
  {\em Lecture Notes in Computer Science}, pages 28--60. Springer, 2009.

\bibitem{CDDFGPZ17}
A.~Cionini, G.~D'Angelo, M.~D'Emidio, D.~Frigioni, K.~Giannakopoulou,
  A.~Paraskevopoulos, and C.~D. Zaroliagis.
\newblock Engineering graph-based models for dynamic timetable information
  systems.
\newblock \emph{Journal of Discrete Algorithms}, Vol.~46-47 (2017), pp.~40-58.

\bibitem{DPW2012}
J.~Dibbelt, T.~Pajor and D.~Wagner.
\newblock User-constrained multi-modal route planning.
\newblock In \emph{Algorithm Engineering and Experiments} -- ALENEX 2012, pp. 118–129. SIAM (2012).

\bibitem{DDPWW2013}
D.~Delling, J.~Dibbelt, T.~Pajor, D.~Wagner, and R.~Werneck.
\newblock Computing multimodal journeys in practice.
\newblock In \emph{12th International Symposium
on Experimental Algorithms (SEA 2013)}, volume 7933 of \emph{Lecture Notes in Computer
Science}, pages 260–271. Springer, 2013.

\bibitem{DPW09}
D.~Delling, T.~Pajor, and D.~Wagner.
\newblock Engineering time-expanded graphs for faster timetable information.
\newblock In {\em Robust and Online Large-Scale Optimization}, volume 5868 of
  {\em Lecture Notes in Computer Science}, pages 182--206. Springer, 2009.

\bibitem{D2016}
J.~Dibbelt.
\newblock Engineering Algorithms for Route Planning in Multimodal Transportation Networks.
\newblock PhD Thesis, Karlsruhe Institute of Technology, February 2016.

\bibitem{FSZ09}
M.~Fischetti, D.~Salvagnin, and A.~Zanette.
\newblock Fast approaches to improve the robustness of a railway timetable.
\newblock {\em Transportation Science}, 43(3):321--335, 2009.

\bibitem{GNPZ2017}
K.~Giannakopoulou, S.~Nikoletseas, A.~Paraskevopoulos, and C.~Zaroliagis.
\newblock Dynamic Timetable Information in Smart Cities.
\newblock in \emph{Proc.~22nd IEEE Symposium on Computers and Communications} -- ISCC 2017,
IEEE Computer Society, pp.~42-47.

\bibitem{GH05}
A.~Goldberg and C.~Harrelson.
\newblock Computing the shortest path: A* search meets graph theory.
\newblock In {\em ACM-SIAM Symposium on Discrete Algorithms (SODA2005)}, pages
  156--165. SIAM, 2005.

\bibitem{G84}
A. Guttman.
\newblock  R-trees: {A} dynamic index structure for spatial searching.
\newblock In Proc.~ACM SIGMOD International Conference on Management of Data (SIGMOD 1984), pp.~47-57.

\bibitem{LSSSP10}
C.~Liebchen, M.~Schachtebeck, A.~Sch{\"o}bel, S.~Stiller, and A.~Prigge.
\newblock Computing delay resistant railway timetables.
\newblock {\em Computers {\&} OR}, 37(5):857--868, 2010.

\bibitem{MMPZ13}
G.~Mali, P.~Michail, A.~Paraskevopoulos, and C.~Zaroliagis.
\newblock A new dynamic graph structure for large-scale transportation
  networks.
\newblock In {\em 8th International Conference on Algorithms and Complexity
  (CIAC2013)}, volume 7878 of {\em Lecture Notes in Computer Science}, pages
  312--323. Springer, 2013.

\bibitem{MS98}
P.~Modesti and A.~Sciomachen.
\newblock A utility measure for finding multiobjective shortest paths in urban multimodal transportation networks. 
\newblock \emph{Eur. J. Oper. Res.} 111(3), 495–508 (1998).

\bibitem{osmfds}
{OpenStreetMap Data Extracts}
\newblock {\url{http://download.geofabrik.de}}.

\bibitem{PSWZ08}
E.~Pyrga, F.~Schulz, D.~Wagner, and C.~Zaroliagis.
\newblock Efficient models for timetable information in public transportation systems.
\newblock {\em ACM Journal of Experimental Algorithmics}, 12(2.4):1--39, 2008.

\bibitem{timeScalRef}
{Reference CPU scores}
\newblock {\url{http://i11www.iti.kit.edu/~pajor/survey}}.

\bibitem{SS10}
M.~Schachtebeck and A.~Sch{\"o}bel.
\newblock To wait or not to wait - and who goes first? delay management with
  priority decisions.
\newblock {\em Transportation Science}, 44(3):307--321, 2010.

\bibitem{trfds}
{Transit Feeds}.
\newblock {\url{https://transitfeeds.com}}.

\bibitem{tfluk}
{Transport for London}.
\newblock {\url{https://tfl.gov.uk}}.

\bibitem{WZ2017}
D.~Wagner and T.~Z\"undorf.
\newblock Public Transit Routing with Unrestricted Walking.
\newblock In \emph{Algorithmic Approaches for Transportation Modeling, Optimization, and Systems} -- ATMOS 2017,
OASIcs Series, Vol.~59~(2017), pp. 7:1-7:14.

\end{thebibliography}

% \newpage

\appendix

\section{Appendix}
\label{sec:app}

Table~\ref{table:queries-alt} shows the performance of the algorithms with (\checkmark) and without ($\times$)
the ALT heuristic for the case of limited walking (the unlimited case exhibited a similar performance
difference).

\begin{table}[htb]
\centering
\small
\hspace*{-2.3cm}
\resizebox{\columnwidth}{!}{%
\begin{tabular}{|c|c|c|c|c|c|c|c|c|c|c|}
\hline
\multirow{2}{*}{~} & \multicolumn{3}{c|}{Algorithm} & \multicolumn{5}{c|}{Travel Modes} & \multicolumn{2}{c|}{\begin{tabular}[c]{@{}c@{}}Query {[}ms{]}\end{tabular}} \\ \cline{2-11}
 & ALT($\times$) & ALT(\checkmark) & MC & Bus & Train & Walk & EV/Car & Cycle & ALT($\times$) & ALT(\checkmark)  \\ \hline
\multirow{4}{*}{\rotatebox{90}{Berlin}} & \redTEEngQuery \cite{CDDFGPZ17} & \redTEQueryALT \cite{CDDFGPZ17} & & $\bullet$ & $\bullet$ & & & & 18.34 & 6.88 \\ \cline{2-11}
 & \dynTMEngQuery \cite{CDDFGPZ17} & \dynTMQueryALT \cite{CDDFGPZ17} & & $\bullet$ & $\bullet$ & & & & 27.63 & 12.17 \\ \cline{2-11}
 & \multiDynTMEngQuery & \multiDynTMQueryALT & & $\bullet$ & $\bullet$ & & & & 14.97 & 6.12 \\ \cline{2-11}
 & \multiDynTMEngQuery & \multiDynTMQueryALT & & $\bullet$ & $\bullet$ & $\bullet$ & $\bullet$ & & 16.03 & 8.49 \\ \hline
 \multirow{6}{*}{\rotatebox{90}{London}} &\redTEEngQuery \cite{CDDFGPZ17} & \redTEQueryALT \cite{CDDFGPZ17} & & $\bullet$ & $\bullet$ & & & & 15.13 & 5.14 \\ \cline{2-11}
 & \dynTMEngQuery \cite{CDDFGPZ17} & \dynTMQueryALT \cite{CDDFGPZ17} & & $\bullet$ & $\bullet$ & & & & 31.12 & 10.25 \\ \cline{2-11}
 & \multiDynTMEngQuery & \multiDynTMQueryALT & & $\bullet$ & $\bullet$ & & & & 6.88 & 4.17 \\ \cline{2-11}
 & \multiDynTMEngQuery & \multiDynTMQueryALT & & $\bullet$ & $\bullet$ & $\bullet$ & $\bullet$ & & 10.14 & 6.10 \\ \cline{2-11}
 & \multiCriModDynTMEngQuery-\texttt{1.0} & \multiCriModDynTMEngQueryALT-\texttt{1.0} & $\bullet$ & $\bullet$ & $\bullet$ & $\bullet$ & $\bullet$ & & 10.26 & 6.29 \\ \cline{2-11}
 & \multiCriModDynTMEngQuery-\texttt{1.2} & \multiCriModDynTMEngQueryALT-\texttt{1.2} & $\bullet$ & $\bullet$ & $\bullet$ & $\bullet$ & $\bullet$ & & 19.04 & 15.44 \\ \hline %%% \cline{2-11}
% & \texttt{MCR-ht} \cite{DDPWW2013,D2016} & & $\bullet$ & $\bullet$ & $\bullet$ & $\bullet$ & $\bullet$ & $\bullet$ & 361.23 & \\ \cline{2-11}
% & \texttt{MR-}$\infty$\texttt{-t10} \cite{DDPWW2013,D2016} & & $\bullet$  & $\bullet$ & $\bullet$ & $\bullet$ & $\bullet$ & $\bullet$ & 21.47 & \\ \hline
\end{tabular}}
\hspace*{-2cm}
\caption{Comparison between query algorithms. Symbol $\times$ (\checkmark) denotes a query algorithm without (with) ALT.
\multiCriModDynTMEngQuery-\texttt{[ALT-]1.0} (\multiCriModDynTMEngQuery-\texttt{[ALT-]1.2}) denotes \multiCriModDynTMEngQuery-\texttt{[ALT]} with $P=1$ ($P=1.2$).
Bullets ($\bullet$) indicate the options taken into account. MC denotes a multicriteria journey
on arrival time and number of transfers.}
\label{table:queries-alt}
\end{table}

\end{document}